# Towards a Property Preserving Transformation from IEC 61131–3 to BIP

Jan Olaf Blech, Anton Hattendorf, Jia Huang

fortiss GmbH, Guerickestraße 25, 80805 München, Germany

November 1, 2018

**Abstract** We report on a transformation from Sequential Function Charts of the IEC 61131–3 standard to BIP. Our presentation features a description of formal syntax and semantics representation of the involved languages and transformation rules. Furthermore, we present a formalism for describing invariants of IEC 61131–3 systems and establish a notion of invariant preservation between the two languages. For a subset of our transformation rules we sketch a proof showing invariant preservation during the transformation of IEC 61131–3 to BIP and vice versa.

This work has been supported in part by the European research project ACROSS under the Grant Agreement ARTEMIS-2009-1-100208.

## 1 Introduction

In this paper we provide a formal definition of a subset of the Sequential Function Chart (SFC) language from the IEC 61131–3 standard and the BIP language. The IEC 61131–3 standard [19] describes languages widely used in the industrial control domain to write programs for programmable logic controllers (PLC). BIP [4] is a language to describe component based systems. It is based on state transitions that are encapsulated into components. These components are connected with each other. Via connectors they can interact and synchronize.

We contribute a transformation specification from SFCs into BIP. This transformation is designed to preserve the behavior of SFCs while using as much of BIPs parallelism as possible.

We also present a description for invariants in the SFC language and show how invariants from BIP can be transformed into SFC invariants and vice versa. A proof that invariant properties discovered on BIP do also hold for the original SFCs is provided to show that the transformations are correct.

## 1.1 Related Approaches

Formal specification and correctness of model to model transformations have been extensively studied in the context of graph-transformations [13]. Our work, however, has been stronger influenced by approaches to guarantee the correctness or distinct properties of compiler runs (e.g., [18, 15, 17]). Our transformation rules for the IEC 61131–3 to BIP transformation can be regarded as the definition for a compiler. Guaranteeing correctness of our transformation by using translation validation like techniques is a long term goal of our work. This could be similar to [6].

The issue of property preservation has been studied in [16]: The work comprises conditions and a proof that under these conditions abstractions preserve temporal logics properties. Our transformations may be regarded as abstractions. The invariant based safety properties studied in this paper are a simple (but yet powerful) case of these properties.

Various other BIP transformations exist that are relevant for our work. Most notably synchronous BIP [11] is a language subset of BIP that is especially suitable for the transformation of synchronous languages into BIP. The languages of IEC 61131–3 are also synchronous. Furthermore the translation from AADL to BIP has been studied [12]. Coq certificates guaranteeing correct invariants for BIP have been studied in [8].

## 1.2 Lifting Safety Properties Through a Transformation Chain

On a general level we are interested in transformation chains where models get transformed into other models and finally into machine code. Some analysis tool might be invoked on one of these representations. In particular we are interested in analyses that discover system invariants on one of the intermediate models. Safety results may be derived from the analysis of such an invariant.

We are interested in lifting such invariants and connected safety results from one representation to another. In particular we are interested in two different use-cases.

**Use-Case 1: Lifting Invariants and Safety Properties to an Earlier Stage in the Model Transformation Chain** This comprises the lifting of invariants and connected safety properties to the beginning of a chain of model to model transformations (cf. Figure 1), i.e., transforming model' to model and then invariant' to invariant. This can be done if the following conditions are fulfilled:

- Models at an earlier stage in the transformation chain allow at least as much behavior as models at the analysis stage in the transformation chain.

- Invariants at the analysis stage are at most as strong as corresponding invariants at earlier stages in the transformation chain.

The fact that invariants may only get stronger when transforming them for an earlier model representation ensures the preservation of safety properties. However, we have to show invariant preservation:

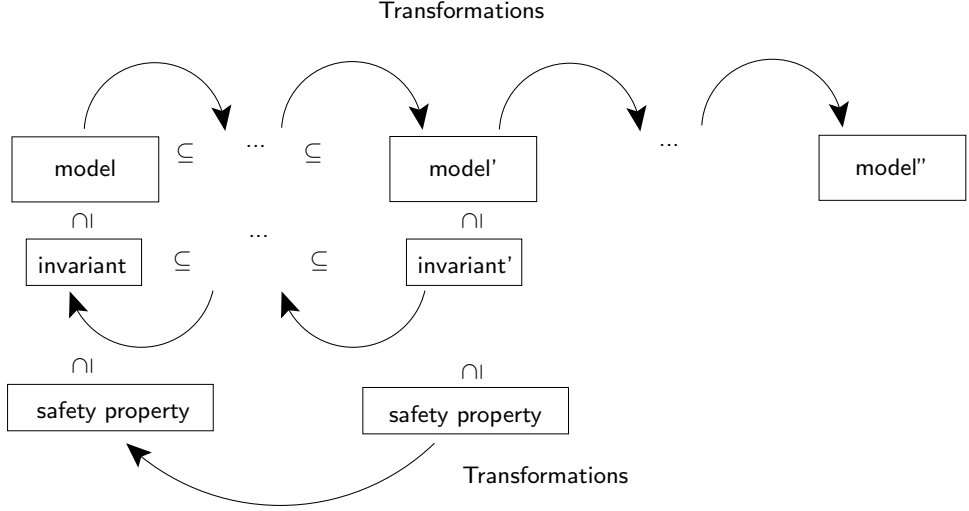

Figure 1: Use-Case 1

- We have to show that for each concrete transformation invariant is indeed an invariant of model.

Correctness of model and invariant transformation can be formally stated in the following way. Given a (model to model) transformation function $T_S$ and an invariant transformation function $T_I$: for every system model there is a transformation model′ = $T_S$(model), if we have discovered an invariant invariant′ for model′ (denoted model′ $\models$ invariant′) then there exists an invariant invariant = $T_I$(invariant′) such that model $\models$ invariant. If $T_i$ preserves safety properties, this ensures that safety properties discovered by analyzing $I_2$ do also hold for the original system model.

**Use-Case 2: Lifting Invariants and Safety Properties to a Later Stage in the Model Transformation Chain** In the second use-case (cf. Figure 2) we start with an invariant invariant and connected safety properties and a model model. All we have to ensure to guarantee the soundness (i.e., safety property preservation) of the approach are the following two items:

- Models at an earlier stage in the transformation chain allow at most as much behavior as models at a later stage in the transformation chain.

- Invariants at an earlier stage are at least as strong as corresponding invariants at later stages in the transformation chain.

In this paper we are regarding transformations from the IEC 61131–3 language to BIP. Invariant-based static analysis tools like D-Finder that work on BIP may be used to discover safety properties like deadlock-freedom. These properties are also valid for the original IEC 61131–3 model if the sketched conditions (first use-case) are fulfilled. Thus, if we have a trustable code-generation from IEC 61131–3 models and do trust D-Finder, we can use this code generation to generate code that we can trust to be deadlock-free given a successful D-Finder run. The other application corresponds to the second use-case: Lifting safety properties and connected invariants from IEC 61131–3 to BIP.

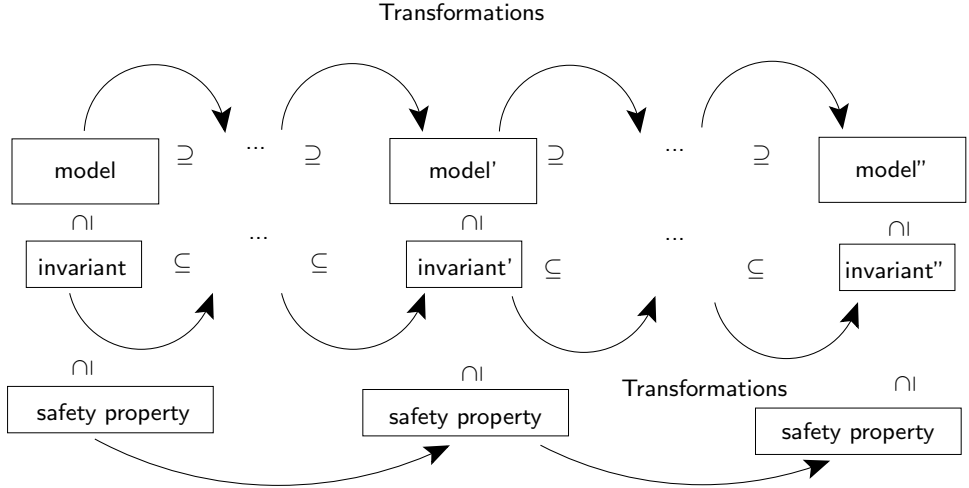

Figure 2: Use-Case 2

### 1.3 Overview

We define a formal semantics of the Sequential Flow Chart part of the IEC 61131–3 standard in Section 2. A summary of the BIP language and its semantics is given in Section 3. The transformation from this language into BIP is described in Section 4. Invariant preservation from BIP back to the SFC language is discussed in Section 5. The other direction: invariant preservation from SFCs to BIP is presented in Section 6. Finally, Section 7 features a conclusion.

## 2 The SFC Language and a Formal Definition of its Semantics

This section presents an overview on the Sequential Function Charts (SFC) language and a formal definition of SFC including the operational semantics. This formulation is based on the IEC 61131-3 standard [19] and existing work [3, 9, 10]. We are particularly interested in the subset of IEC 61131–3 that is used in the EasyLab toolkit [2].

### 2.1 A Formal Definition of the SFC Syntax

SFC is one of the graphical programming languages described in the IEC 61131-3 standard. It comprises control locations of the system (called *steps*) and the transition of control between them. The passing of control can be restricted via *guards*. Behaviors of the program is described in *action* blocks which can be associated with steps. In the following, we first define the basic component of SFCs and then describe the composition of them. We define in fact two formal descriptions of the SFC language. One that supports the subset used by the Easylab toolkit and an extended version that supports more features described in the standard.

**Variables in the SFC Language** SFCs have variables that are visible to all their components, such as steps, guards and actions. We use $X = \{x_1, x_2, ..., x_n\}$ to denote the set of variables. The current values of $X$ are described using a variable valuation function (usually denoted $f$ in the context of this paper) of type $X \to val_X$, which assigns a value compatible to the respective data type ($val_X$) to each variable in $X$. We use $\mathcal{F}$ to denote the set of all valuation functions of $X$.

**Action Blocks and Steps** (Extended) action blocks and steps in SFCs are the basic units for describing the behavior.

**Definition 1 (Actions)** *An action is an update function of type $\mathcal{F} \to \mathcal{F}$.*

The definition of actions is the same for extended and non-extended SFCs.

**Definition 2 ((Extended) Action block)** *For a given set of actions A an extended action block is a tuple $(a, q)$, where*
*– a is the action, which can be either an update function of type $\mathcal{F} \to \mathcal{F} \in A$ or an identifier to a nested SFC, and*
*– q is a qualifier $q \in \{N, S, R, P_0, P_1\}$, the semantics of which are introduced shortly.*

*A non-extended action block comprises just an update function of type $\mathcal{F} \to \mathcal{F}$. In this case, the term action block and update function may be used synonymously.*

The same action can be used in multiple action blocks associated with different SFC steps. In IEC 61131–3 the update functions can be written in any other languages defined in the standard, such as Function Block Diagram (FBD), Structured Text (ST), Ladder Diagram (LD) and Instruction List (IL). Here we abstract from a concrete language and use mathematical functions instead.

The action block qualifiers describe the activity of actions. The semantics of the qualifiers are defined as follows:

- $N$ (non-stored) qualifier describe actions that are active only when parent steps are activated,

- $S$ (stored) qualifier describe actions that continue being active until a reset action is executed,

- $R$ (stored) actions are used for reseting active actions,

- $P_0$ (pulse at falling edge) actions are active only when the parent step is exited,

- $P_1$ (pulse at rising edge) actions are active only when the parent step are entered.

The reset actions tagged with $R$ have always the highest priority.

**Definition 3 (Step)** *For a given set of action blocks $B$ and its associated set of actions $A$. A step of an SFC is a pair $s = (n, \Omega)$, where $n$ is a unique identifer for the step and $\Omega = 2^B$ (note that $B = A$ for the non-extended case) is a set of action blocks belonging to the step. We use $s.\Omega$ to refer to the set of action blocks associated with step $s$.*

The steps can be in *inactive, ready* or *active* state. An inactive step does not activate any actions. The ready steps are the ones that control resides in. However, it is not necessary that the actions of ready steps will be performed. A special situation is in nested SFCs, where a step in a nested SFC is ready but the parent action is not active. In this case, the actions of ready steps will not be performed. The active steps are those ready steps whose actions will actually be performed.

**Guards and Transitions**  Steps in SFCs are connected via transitions. Transitions feature a guard expression.

**Definition 4 (Guard)** *A guard $g$ is a predicate over a valuation function. It has the type $g : \mathcal{F} \to bool$, where $F$ is the set of all valuation functions. It evaluates to true if the current values of $X$ satisfy $g$.*

**Definition 5 (Transition)** *A transition $t \in T = (t_{src}, t_g, t_{tgt})$ describes the moving of control from source steps $t_{src}$ to target steps $t_{tgt}$. A transition is enabled if the guard $t_g$ evaluates to true. A transition is taken if no conflicting transition with higher priority is enabled.*

**SFC Definition**  The definition of SFCs is the same for both extended and non-extended SFCs. However, the set of actions has a different type.

**Definition 6 (SFC)** *An SFC is a 7-tuple $\mathcal{S} = (X, A, S, S_0, T, \sqsubset, \prec)$, where*

- *$X$ is a finite set of variables,*

- *$A$ a finite set of actions,*

- *$S$ is a finite set of steps, comprising action blocks which depend on $A$ (in case of non-extended SFCs are equal to $A$)*

- *$S_0$ is the set of initial steps,*

- *$T \subseteq (2^S \setminus \{\emptyset\}) \times G \times (2^S \setminus \{\emptyset\})$ is the set of transitions, where $G$ is the set of guards,*

- *$\sqsubset \in A \times A$ is a total order on actions used to define the order in which the active actions are to be executed, and*

- *$\prec \in T \times T$ is a partial order on transitions to determine the priority on conflicting transitions.*

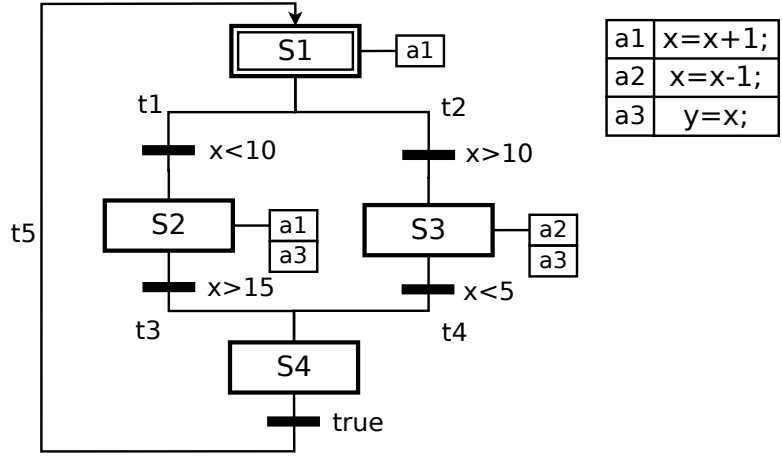

Figure 3: An example SFC

**Example**  Figure 3 depicts an example non-extended SFC model consisting of four steps and three actions. Its formal definition is:

$$\begin{aligned}
\mathcal{S} = \{&\{x,y\},\\
&\{S1, S2, S3, S4\},\\
&\{a1, a2, a3\},\\
&\{S1\},\\
&\{\ (\{S1\}, x1 < 10, \{S2\}), (\{S1\}, x > 10, \{S3\}), (\{S2\}, x > 15, \{S4\}),\\
&\quad (\{S3\}, x < 5, \{S4\}), (\{S4\}, true, \{S1\}),\},\\
&\{a1 \sqsubset a2 \sqsubset a3\},\\
&\{t1 \prec t2\},\\
\}&
\end{aligned}$$

### 2.2 Operational Semantics of SFCs

As proposed in [3], the operational semantics for SFCs is based on configurations describing the system state [1]:

**Definition 7 ((extended) Configuration)** *An extended configuration of an SFC and its sub-SFCs is a 5-tuple $c = (f, readyS, activeS, activeA, storedA)$, where $f$ is the function describing the current values of variables. The rest are four sets describing the states of steps and actions. In particular, readyS is the set of ready steps, activeS is the set of active steps, activeA is the set of active actions and storedA is the set of stored actions. Non extended configurations have the following form: $(f, activeS, activeA)$.*

---

[1] In this paper, we use the term system state and configuration interchangeably

**Semantics of Extended SFCs** Execution of one SFC-cycle consists of the following three phases:

- Phase 1: execute actions contained in set $activeA$ and update the values of variables correspondingly;

- Phase 2: perform step transitions and update the sets $readyS$, $activeS$;

- Phase 3: compute the set of active/stored actions $activeA$ and $storedA$ for the next cycle.

In each of the three phase, the configuration of the SFC system is updated. The semantics of an SFC can then be regarded as a transition system of the configurations.

**Definition 8 (Transition system of an extended SFC)** *An extended SFC*
$$\mathcal{S} = (X, A, S, S_0, T, \sqsubset, \prec)$$
*is associated with a transition system $\mathcal{E}(S) = (\mathcal{C}, c_0, \rightarrow)$, where $\mathcal{C}$ is the set of configurations, $c_0$ is the initial configuration and $\rightarrow \subseteq C \times C$ is the transition relation.*

Elements in the transition relation for extended SFCs have the following form:

$(f, readyS, activeS, activeA, storedA) \rightarrow (f', readyS', activeS', activeA', storedA')$

The details of the transition relation are described as follows:

- In Phase 1, $f'$ is updated by performing the set of active actions, which are executed according to the orders specified in $\sqsubset$, that is, $f' = (a_m \circ ... \circ a_1)(f)$, where $a_1 \sqsubset ... \sqsubset a_m$ and $\{a_1, ..., a_m\}$ are the set of active update functions. The rest of the configuration remains unmodified.

- In Phase 2, $readyS$, $activeS$ are updated and the rest of the configuration remains the same. $readyS'$ is obtained by evaluating transitions $T$. Let $c$ be the current configuration, the set of transitions that can be taken can be identified as $\mathcal{T} = \{t = (t_{src}, t_g, t_{tgt}) \in T \mid t_{src} \subseteq activeS \land c \models t_g \land (\neg \exists t' . t'_{src} \subseteq activeS \land c \models g' \land t_{src} \cap t'_{src} \neq \emptyset \land t \prec t')\}$, that is, a transition is taken if its guard is satisfied and the conflicting transitions, which are transitions originated from the same step, are either not enabled or of lower priority. Then, the active steps for the next cycle can be obtained as $readyS' = readyS \setminus \{t_{src} | t \in \mathcal{T}\} \cup \{t_{tgt} | t \in \mathcal{T}\}$. Since actions can be associated with multiple steps, it is also possible that actions get activated multiple times with different qualifiers. Hence, to decide if a step or action is in active state, we need to inspect all its activations. For that, the iterative methods proposed in [3] can be used. From the top level SFC, the algorithm iterates over all nested SFCs to find all active steps ($activeS'$).

- In Phase 3, $activeA$ and $storedA$ are updated and the rest of the configuration remains the same. The iterative algorithm in [3] finds beside $activeS'$ also the set of all qualifiers that appear in activations of each action. We use $\alpha(a)$ to denote this set of qualifiers for action $a$. Then:
  - $activeA' = \{a \in A \mid \alpha(a) \cap \{N, P_0, P_1\} \neq \emptyset \land R \notin \alpha(a)\}$
  - $storedA' = \{a \in A \mid S \in \alpha(a) \land R \notin \alpha(a)\}$

**Semantics of Non-extended SFCs** The focus of this paper is to investigate invariant preservation properties for a subset of SFCs supported by EasyLab (non-extended SFCs), in which the following restrictions are added:
– the nested SFCs are not allowed;
– only $N$ qualifier is supported.
With these restriction, the configurations can be simplified by eliminating the sets $readyS$ and $storedA$, because $readyS = activeS$ always holds and $storedA$ is always empty. The case of the entire SFC language will be addressed in future work.

**Definition 9 (Transition Systems of non-extended SFCs)** *An non-extended SFC is also represented as a tuple $\mathcal{S} = (X, A, S, S_0, T, \sqsubset, \prec)$. Due to the simplification, the transition relation has the form:*

$$(f, activeS, activeA) \to (f', activeS', activeA')$$

For non-extended SFCs, a simplified version of operational semantics can be defined and the set of reachable configurations can be regarded. Let $[\![SM]\!]_{SFC}$ denote the set of all possible configuration transitions of a non-extended SFC $SM$. It can be formally defined as follows:

$$(c, c') = ((f, activeS, activeA), (f', activeS', activeA')) \in [\![SM]\!]_{SFC}$$

$$iff$$

$$executeAction(c, c') \vee stepTransition(c, c') \vee activateAction(c, c')$$

The term $executeAction$, $stepTransition$ and $activateAction$ represent the three possible types of configuration transitions. Formally,

$$executeAction(c, c') = \exists\ \hat{a} \in activeA\ .\ f' = \hat{a}(f)$$
$$\wedge\ activeS = activeS' \wedge activeA' = activeA \backslash \{\hat{a}\}$$
$$stepTransition(c, c') = \exists\ t \in T\ .\ t_{src} \subseteq activeS \wedge t_g(f) \wedge f = f'$$
$$\wedge\ (activeS' = activeS \setminus t_{src} \cup t_{tgt}) \wedge activeA = activeA'$$
$$activateAction(c, c') = \exists\ s \in activeS.\exists a \notin activeA.a \in s.\Omega$$
$$\wedge\ f = f' \wedge activeS = activeS' \wedge activeA' = activeA \cup \{a\}$$

The reachable configurations of $SM$ can then be inductively defined as follows, demanding that the initial state is reachable and each reachable configuration must be able to be reached from it via valid transitions.

$ReachableConfig_{SFC}(c) =$

$\left. \begin{array}{l} \quad c = c_0 \\ \vee\ \exists c'.ReachableConfig_{SFC}(c') \wedge (c', c) \in [\![SM]\!]_{SFC} \end{array} \right\}\ smallest\ fixpoint$

The elements in $[\![SM]\!]_{SFC}$ are the basic configuration transition steps of $SM$. For the rest of the paper, we introduce the notation $(c \xrightarrow{SM}_{SFC} c')$, which means $(c, c') \in$ $[\![SM]\!]_{SFC}$, i.e., $c'$ can be reached from $c$ via a single transition step contained in $[\![SM]\!]_{SFC}$. Another notation ($c \xrightarrow{\text{SM}}{}^+_{SFC} c'$) means $c'$ can be reached from $c$ via multiple of the transition steps contained in $[\![SM]\!]_{SFC}$.

## 3 BIP and its Semantics

In this section we present a subset of the BIP language. BIP allows the modeling of asynchronous components and of interactions between them. We discuss its semantics and an example BIP model. Parts of this section follow the presentation given in [7] building upon [4].

### 3.1 BIP models

BIP (Behavior, Interaction, Priority) is a software framework designed for building embedded systems consisting of asynchronously interacting components, each specified as a non-deterministic state transition system. Tools developed for BIP comprise static analyzers and code generation.

**Atomic Components** BIP models are composed of atomic components [4, 5]. An atomic component $(L, P, T, V, D)$ is a state transition system consisting of a set of locations $L$, a set of ports $P$, a set of transitions $T$, and a set of variables $V$ which are mapped to values of type $D$. An atomic component has a distinct state of type

$$L \times (V \to D)$$

comprising a location and a variable valuation. The latter is a mapping from variables to their values. Transitions are of type

$$T \subset L \times ((V \to D) \to bool) \times ((V \to D) \to (V \to D)) \times P \times L$$

They comprise a source location, a guard function, an update function, a port, and a target location. Our semantics requires that a transition from one location to another can be performed iff the guard function evaluates to true using the variable valuation in the current state. During a transition the variable valuation is updated for the succeeding state. Furthermore, it is possible to restrict transitions by putting constraints on the port involved in the interaction.

Each port $p \in P$ can have an associated variable. This variable is used to exchange data between different atomic components in composed components (see below). The functions $\mathcal{V}(p) : P \to V \cup \{\epsilon\}$ defines the association between port and variable. If the result of $\mathcal{V}$ is $\epsilon$, the port does not exchange data.

Figure 4 shows an atomic component comprising two locations ($l_5$ and $l_6$), and a variable $\theta$ that has a numeric value. Four possible transitions exist between these locations which are depicted as arcs. Each one comprises a port (*tick*, *cool*, *heat*), a guard which checks for conditions of the variable $\theta$, and an update functions which is empty on two arcs ($--$) and modifies the value of the variable $\theta$ on the other arcs.

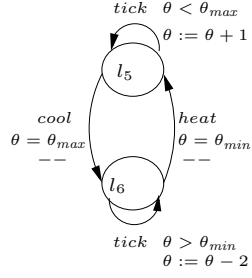

Figure 4: An Atomic Component

**Composed Components** Atomic components may be glued together to form composed components. The behavior of the resulting system can be restricted by linking components with connectors. These put constraints on ports in the different atomic components. A composed component is a tuple $(A, C)$ comprising a set of atomic components $A$ and a set of connectors $C$.

$$C \subset (A \times P) \times 2^{A \times P}$$

Connectors have the following form: $((a_s, p_s), \{(a_1, p_1), (a_2, p_2), ..., (a_n, P_n)\})$ comprising an atomic component $a_x \in A; x \in \{s, 1...n\}$ and a relevant port $p_x \in P_{a_x}$ with $P_{a_x}$ being the set of ports associated with $a_x$. If the connector is used to exchange data, the data is copied from $a_s$ to all $a_i; i \in 1...n$. For connectors without data exchange there is no difference for $a_s$ and $a_i$.

In an extended version the connectors may contain guard functions, depending on the variable valuations of the linked components, and update functions which are performed on these variable valuations and mechanisms for value exchange between components.

Gluing components together by using connectors realizes weak and strong synchronizations as well as broadcasts between components. Update functions on the involved variable valuations are used to pass values between components.

**Semantics of Composed Components** The state of a composed component is the product of its atomic components' states: $(L_1 \times (V_1 \to D_1)) \times ... \times (L_m \times (V_m \to D_m))$. A state transition relation $[\![BM]\!]_{BIP}$ is defined upon them for a composed component $BM = (A, C)$. We assume an indexing of atomic components $A = \{a_1, ..., a_m\}$

$$(((l_1, x_1), ..., (l_m, x_m)) \,,\, ((l'_1, x'_1), ..., (l'_m, x'_m))) \in [\![BM]\!]_{BIP}$$

iff

$$
\begin{aligned}
\exists (c_s, C_r) \in C : \forall i \in \{1...m\} : \\
(\exists j \in \{1...m\} : \\
(\exists p : (a_j, p) = q_s \wedge \mathcal{V}(p) = v_j) \\
\wedge \mathcal{V}(p_i) = v_i \\
\wedge (l_i, g_i, f_i, p_i, l'_i) \in a_i.T \\
\wedge g_i(x_i) \\
\wedge x'_i = f_i(x_i)[v_i \leftarrow x_j(v_j)] \\
\wedge (a_i, p_i) \in (\{c_s\} \cup C_r)) \vee \\
(l_i = l'_i \\
\wedge \neg(\exists p : (a_i, p) \in (\{c_s\} \cup C_r)) \\
\wedge x_i = x'_i)
\end{aligned}
$$

$a_i.T$ denotes the set of transitions associated with component $a_i$

**Reachable States of BIP models**  The set of reachable states for a BIP model $BM$ for a given initial state $s_0$ is defined by a predicate $R_{BM}$ via the following inductive rules:

$$
\frac{}{R_{BM}(s_0)} \qquad \frac{R_{BM}(s) \quad (s, s') \in [\![BM]\!]_{\mathsf{BIP}}}{R_{BM}(s')}
$$

The first rule says that the initial state is reachable. The second inference rule captures the transition behavior of BIP using the transition relation.

### 3.2 An Example

Figure 5 shows a temperature control system modeled in BIP, which has been well discussed in the literature (e.g., [5, 1, 14]). It uses the component from Figure 4 but puts constraints on the ports in order to use it as a controller.

The system controls the cooling of a reactor by moving two independent control rods. Each one has its own timer $t_1$, $t_2$. After the usage of a rod there is a timeout $t_{max}$ until it can be reused again. The goal is to keep the temperature $\theta$ between $\theta_{min}$ and $\theta_{max}$. When the temperature reaches the maximum value, one of the rods has to be used for cooling. The BIP model comprises three atomic components: one for each rod and one for the controller. Each contains a state transition system. Transitions can be labeled with guard conditions, valuation function updates, and a port. The components interact via ports thereby realizing cooling, heating, and time elapsing interactions. In the figure, possible interactions are indicated by arcs connecting ports from different components. These require that either in every connected component a transition labeled with the connected port must be taken in parallel or none of these transitions is taken. For example, the time elapsing *tick* transitions must be taken in each component in the same parallel step (*tick1*, *tick*, and *tick2*). Depending on the values of $\theta_{max}$, $\theta_{min}$, and $t_{max}$ the system might either contain a deadlock or not.

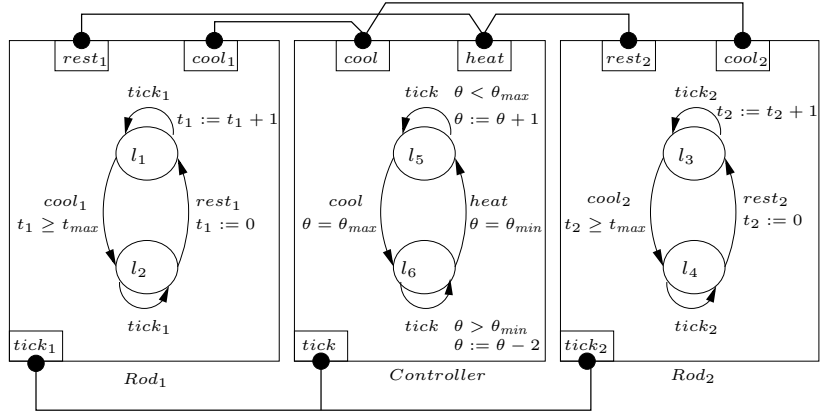

Figure 5: Temperature Control System

Let us introduce two invariants for the example model with $\theta_{max} = 1000$, $\theta_{min} = 100$, and $t_{max} = 3600$:

1. $I_1 \equiv (at_{l_5} \wedge 100 \leq \theta \leq 1000) \vee (at_{l_6} \wedge 100 \leq \theta \leq 1000)$
   This invariant states that the temperature will always be between 100 and 1000.

2. $I_2 \equiv (at_{l_1} \wedge t_1 = 0) \vee (at_{l_3} \wedge t_2 = 0) \vee (at_{l_5} \wedge 101 \leq \theta \leq 1000) \vee$
   $(at_{l_6} \wedge (\theta = 1000 \vee 100 \leq \theta \leq 998))$
   This invariant states that the timer in one of the rods is zero or we are in the heating phase and the temperature is already above 100 or we are in the cooling phase and the temperature is 1000 or between 100 and 998.

$at_i$ is a predicate denoting the fact that we are at location $i$ in a component. The first invariant is restricted to a single component, the second one combines facts about several components.

## 4 Transformation from SFCs to BIP

This sections describes the transformation rules from the SFC language to BIP.

For a given SFC $\mathcal{S} = (X, A, S, S_0, T, \sqsubset, \prec, h)$, the transformed BIP model can be represented by a composed component $\mathcal{B} = (\hat{A}, \hat{C})$, with $\hat{A} \subset L \times P \times T \times V \times D$ being the set of atomic components and $\hat{C} \subset (\hat{A} \times P) \times 2^{\hat{A} \times P}$ the set of connectors. In the following sections, we first describes the atomic components including their formal definition (4.1), and afterward the transformations steps and connections of atomic components (4.2).

**SFC evaluation phases** As introduced in Section 2.2, execution of an SFC is done in SFC cycles, which consists of three phases:

1. execution of active actions,

2. evaluation of transitions and identification of active steps,

3. identification of active actions for the next cycle.

In the generated BIP model, an *SFCManager* component is introduced to enforce synchronization of the phases. This is done with the $wTick$ (work tick), $tTick$ (transition tick) and $fTick$ (finish tick) signals.

## 4.1 Atomic Components

Several transformation templates can be identified for the transformation of SFC elements into BIP elements.

**Actions**  Each SFC action $a \in A$ will be represented by a BIP component $\hat{a} \in \hat{A}$. Each action has two ports for synchronization with the BIP representation of the SFC program. The *work* port starts the action. Then it reads all input data, processes the data and writes the output value. At last is enables the *done* port to signal its ACB (see next paragraph) that it has finished processing.

To access global variables, the action has a $read_x$ and/or a $write_x$ for each variable $x$ is accesses. They are connected to the *read* and *write* ports of the global variable atomics.

**Action Control Blocks**  For an action component $\hat{a} \in \hat{A}$ transformed from an SFC action $a \in A$, a special atomic component called Action Control Block (ACB) is equipped, which is responsible for collect and evaluate qualifiers from action activations to decide if the action will be executed for the next SFC cycle. We use $\hat{b}_a$ to denote the ACB component created for $\hat{a}$ and $\hat{B}$ for the set of ACB components for all actions.

**Transformation 1 (Action Control Block)**

$\text{CreateACB}(a) : A \to L \times P \times T \times L \times V \times D$

$\text{CreateACB}(a) = (L_{ACB}, P_{ACB}, T_{ACB}, L_{0_{ACB}}, V_{ACB}, D_{ACB})$

*with the following substitutions:*

$$L_{ACB} = \{ACTIVE_a, WORK_a, WORKED_a, WAIT_a\}$$
$$P_{ACB} = \{wTick_a, tTick_a, N_a, S_a, R_a, work_a, done_a\}$$
$$T_{ACB} = \{$$
$$(ACTIVE_a, true, \{\}, tTick_a, WAIT_a),$$
$$(WAIT_a, true, \{e \leftarrow (\neg r \wedge (s \vee n)); s \leftarrow (\neg r \wedge s); r \leftarrow 0; n \leftarrow 0\},$$
$$wTick_a, ACTIVE_a),$$
$$(ACTIVE_a, e, \{\}, work_a, WORK_a),$$
$$(WORK_a, true, \{\}, done_a, WORKED_a),$$
$$(WORKED_a, true, \{\}, tTick_a, WAIT_a),$$
$$(WAIT_a, true, \{n \leftarrow 1\}, N_a; WAIT_a),$$
$$(WAIT_a, true, \{r \leftarrow 1\}, R_a; WAIT_a),$$
$$(WAIT_a, true, \{s \leftarrow 1\}, S_a; WAIT_a),$$
$$\}$$
$$L_{0_{ACB}} = \{WAIT_a\}$$
$$V_{ACB} = \{n_a : bool, s_a : bool, r_a : bool, e_a : bool\}$$
$$D_{ACB} = \{0, 1\}$$

A graphical representation can be found in Figure 6. Each ACB contains a set of boolean variables $r$, $n$ and $s$, which are set to true if the respective qualifiers $R$, $N$ and $S$ appeared in the previous SFC cycle. At the end of the previous SFC cycle, the collected qualifiers are evaluated to determine if the action will be executed (by computing $e$) and if it will be stored (by computing $s$). In the beginning of the current SFC cycle, the ACB is triggered to $ACTIVE$ by $wTick$ signal. Then, the associated action is executed by sending the $work$ signal if the boolean variable $e$ computed from the previous cycle is true. When the execution of all actions finishes, the SFC manager triggers all ACBs to the $WAIT$ location by $tTick$ signal, in which action activations for next SFC cycle are collected. The $fTick$ signal triggers the ACBs to back $IDLE$ locations at the end of current SFC cycle.

For non-extended SFCs, the ACB component can be simplified. Unused variables $r$, $s$ and $n$ can be eliminated and gathering of quantifiers can be modeled as a simple location transition. Figure 7 is such a simplified guard component.

**Steps** A step in SFC $s \in S$ is represented by an Atomic BIP Component $\hat{s} \in \hat{S}$ defined as follows.

**Transformation 2 (Step)**

$$\text{CreateStep}(s) : S \to L \times P \times T \times L \times V \times D$$
$$\text{CreateStep}(s) = (L_{Step}, P_{Step}, T_{Step}, L_{0_{Step}}, \{\}, \{\})$$

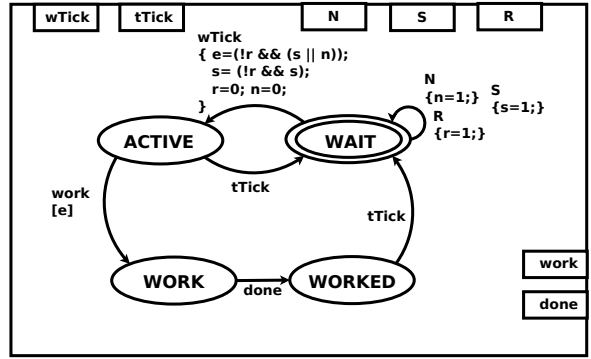

Figure 6: BIP Atomic Component for an Action Control Block

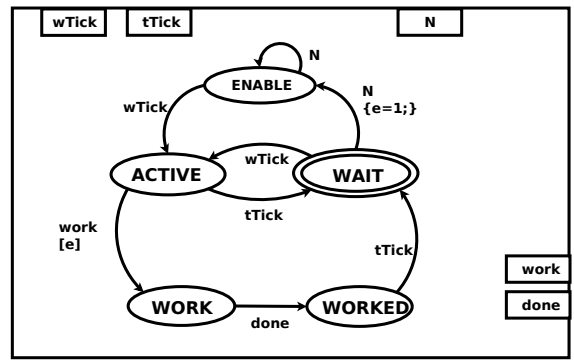

Figure 7: BIP Atomic Component for an Action Control Block for Non-extended SFCs

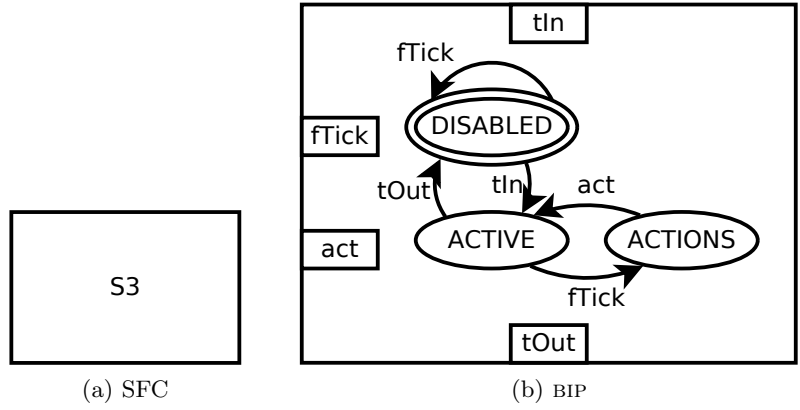

Figure 8: Transformation of an SFC Step into BIP

*with the following substitutions:*

$$L_{Step} = \{DISABLED_s, ACTIVE_s, ACTION_s\}$$
$$P_{Step} = \{tIn_s, tOut_s, fTick_s, act_s\}$$
$$T_{Step} = \{$$
$$(ACTIVE_s, true, \{\}, tOut_s, DISABLED_s),$$
$$(DISABLED_s, true, \{\}, tIn_s, ACTIVE_s),$$
$$(DISABLED_s, true, \{\}, fTick_s, DISABLED_s),$$
$$(ACTIVE_s, true, \{\}, fTick_s, ACTION_s),$$
$$(ACTION_s, true, \{\}, act_s, ACTIVE_s),$$
$$\}$$
$$L_{0_{Step}} = \{DISABLED_s\}$$

A graphical representation can be found in Figure 8b.

The *tIn* and *tOut* Ports are used to enable and disable the steps; *fTick* is used for synchronization within the SFC evaluation phases. The *act* port is used to activate actions.

After the *tTick* has been sent by the SFC manager, the step transition guards (introduced later) will evaluate all transition conditions. For an enabled transition, the BIP components representing the set of source steps will be forced from $ACTIVE$ to $DISABLED$ by receiving the *tOut* signal from the guard. Similarly, the set of BIP components representing the target steps are triggered to $ACTIVE$ location. Further on, on receiving the *fTick*, activation signals are sent to ACBs of actions associated with active steps.

**Guard** For each step transition $t = (t_{src}, t_g, t_{tgt}) \in T$ in SFC, a guard $t_g$ is associated. This guard is built as an atomic component $\hat{t}_g \in \hat{A}_G$ in the BIP domain. The definition of the guard atomic component is as follows:

**Transformation 3 (Guard)**

$$\text{CreateGuard}(t) : T \to L \times P \times T \times L \times V \times D$$
$$\text{CreateGuard}(t) = (L_{Guard}, P_{Guard}, T_{Guard}, L_{0_{Guard}}, \{\}, \{\})$$

*with the following substitutions:*

$$L_{Guard} = \{WAIT_t, READ_t, GUARD_t, ACT_t, DONE_t\}$$
$$P_{Guard} = \{val_t, guard_t, tTick_t, fTick_t, act_t\}$$
$$T_{Guard} = \{$$
$$\quad (WAIT_t, true, \{\}, tTick_t, READ_t),$$
$$\quad (READ_t, true, \{\}, val_t, GUARD_t),$$
$$\quad (GUARD_t, true, \{\}, fTick_t, WAIT_t),$$
$$\quad (GUARD_t, g, \{\}, guard_t, ACT_t),$$
$$\quad (ACT_t, g, \{\}, act_t, DONE_t),$$
$$\quad (DONE_t, true, \{\}, fTick_t, WAIT_t)$$
$$\}$$
$$L_{0_{Guard}} = \{WAIT_t\}$$

The most important part of the guard is $g$, which represents the condition for this guard. Because the conditions may differ, a BIP guard component has to be customized for each SFC guard. Also, access to variables has to be added. The sample in Transformation 3 just reads one variable $val$. Figure 9b contains a graphical representation of the sample guard.

In $WAIT$ location the guard is waiting for begin of the second phase (port $tTick$), in which it first reads the global variables that are necessary to evaluate $g$. Then, if the condition $g$ evaluates to true, the $guard$ port is activated. If the transition the guard is connected to can be taken, the guard component moves to $ACT$. The $guard$ port is connected to the $tIn/tOut$ Ports of the BIP components that represent predecessor/successor SFC steps. The location $ACT$ is introduced to activate actions with $P0$ and $P1$ qualifiers, respectively. The activation ports are connected to the $N$ ports of the associated actions. Note that for both $P0$ and $P1$ activations, the activation ports of a guard send actually a $N$ activation. This is because the behavior of $P0/P1$ activation is the same as that of $N$ activations, except that $P0/P1$ activations are only enabled during step transitions. The second phase of an SFC cycle is finished with $fTick$. In Figure 9b a graphical representation of a guard component is presented.

For non-extended SFCs, the qualifier $P0$ and $P1$ is not used and associated activation port can be eliminated. Figure 10 shows a BIP guard component for non-extended SFCs.

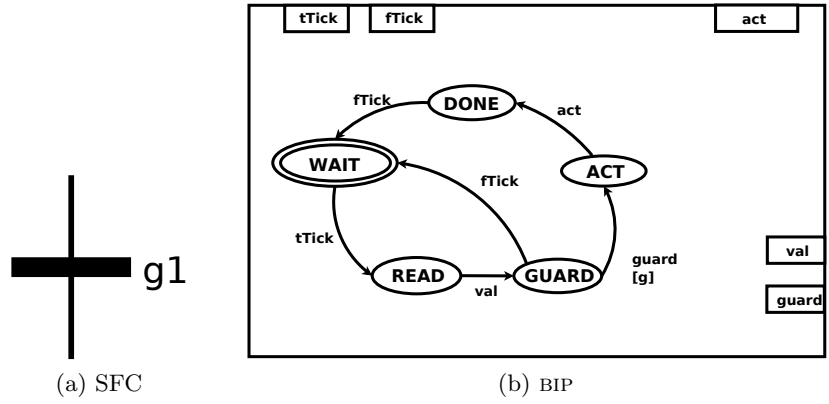

(a) SFC   (b) BIP

Figure 9: Transformation of a Guard Component

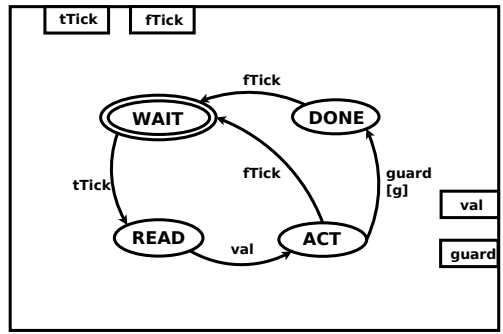

Figure 10: The BIP Guard Component for non-extended SFC

**SFC Manager** The SFC Manager component, displayed in Figure 11, enforces the synchronization of the SFC execution phases by periodically enabling its $wTick$, $tTick$ and $fTick$ ports:

**Transformation 4 (SFC Manager)**

$$\hat{m} = (L_M, P_M, T_M, L_{0_M}, \{\}, \{\})$$
$$L_M = \{WORK_m, TRAN_m, DONE_m\}$$
$$P_M = \{wTick_m, tTick_m, fTick_m\}$$
$$T_M = \{$$
$$\qquad (WORK_m, true, \{\}, tTick_m, TRAN_m),$$
$$\qquad (TRAN_m, true, \{\}, fTick_m, DONE_m),$$
$$\qquad (DONE_m, true, \{\}, wTick_m, WORK_m)$$
$$\qquad \}$$
$$L_{0_M} = \{DONE_m\}$$

Synchronization is done by connecting all $wTick$, $tTick$, $fTick$ ports of all components that need the signal.

The three phases are:

- WORK: This phase is started by the $wTick$. During this phase all enabled actions are executed by their action control blocks. During this phase the actions read global variables (including input data), process them and write the results back. The results don't get valid in this phase.

- TRAN: This phase starts with the $tTick$. With the $tTick$ the results of the actions get valid. Afterwards the conditions of the guards are evaluated and, if possible, taken. SFC Actions that are called with the $P0$ or $P1$ action qualifiers are marked in their ACB during this phase.

- DONE: The third phase is initiated by the $fTick$, after all possible transitions have been taken. In this phase all actions that should be executed in the next round are determined by evaluating the action qualifiers. The ACB stores this information.

**Variables** The BIP language does not support global variables that are accessible to all components. Hence, an SFC variable $x \in X$ is encapsulated as an atomic BIP component Global Variable (GV) defined in the following:

**Transformation 5 (Global Variable)**

$$\text{CreateGV}(x) : X \to L \times P \times T \times L \times V \times D$$
$$\text{CreateGV}(x) = (L_{Global}, P_{Global}, T_{Global}, L_{0_{Global}}, V_{Global}, D_{Global})$$

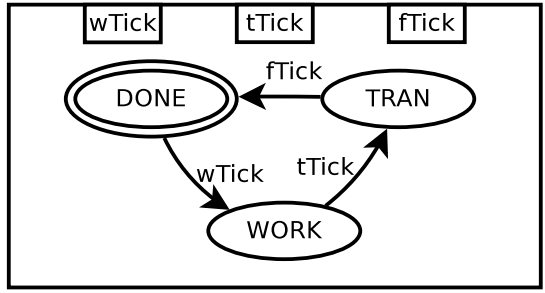

Figure 11: SFC Manager

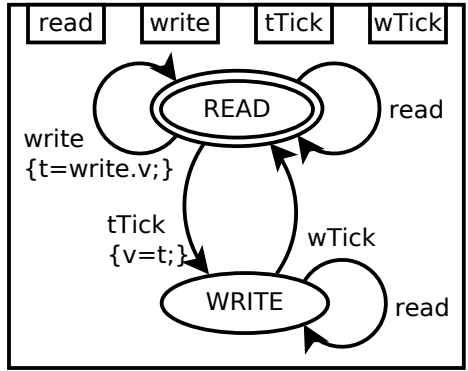

Figure 12: Global Variable

*with the following substitutions:*

$$\begin{aligned}
L_{Global} &= \{READ_x, WRITE_x\} \\
P_{Global} &= \{read_x, write_x, tTick_x, fTick_x\} \\
T_{Global} &= \{ \\
&\quad (READ_x, true, \{\}, read_x, READ_x), \\
&\quad (READ_x, true, \{t_x = write_x.v\}, write_x, READ_x) \\
&\quad (READ_x, true, \{v_x = t_x\}, tTick_x, WRITE_x) \\
&\quad (WRITE_x, true, \{\}, read_x, WRITE_x), \\
&\quad (WRITE_x, true, \{\}, wTick_x, READ_x) \\
&\quad \} \\
L_{0\,Global} &= \{READ_x\} \\
V_{Global} &= \{t_x, v_x\} \\
D_{Global} &= \mathbb{Z}
\end{aligned}$$

The GV component contains two variables $t$ and $v$, both are of the same type as the corresponding SFC variable. The variable $v$ hold the actual value of the SFC variable whereas $t$ is a temporary buffer. Initially, the GV component is in $READ$ location, in which reading of the variable value is allowed and writing of the variable is buffered in $t$. The buffer value is written to $v$ on receiving the $tTick$ signal, that is, when all actions has been performed. This guarantees that all actions read the consistent value of $v$. On the other hand, the buffered value is written immediately with the $tTick$, so that the guard components can read the up-to-date value of variables and evaluate the transition conditions correctly. The $wTick$ brings the GV back to $WRITE$ location. Figure 12 is the graphical representation of the GV component, in which $write.v$ is the value exchanged via the write port.

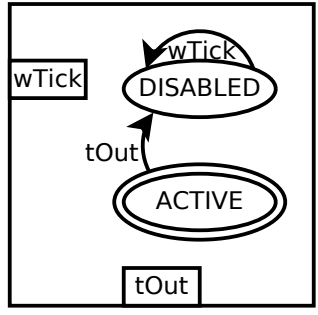

Figure 13: SFC Starter

**Starter** This component activates the initial steps of the SFC program:

**Transformation 6 (Starter)**

$$\text{CreateStarter}(r) : S_0 \to L \times P \times T \times L \times V \times D$$
$$\text{CreateStarter}(r) = (L_{Starter}, P_{Starter}, T_{Starter}, L_{0_{Starter}}, \{\}, \{\})$$

*with the following substitutions:*

$$L_{Starter} = \{DISABLED_r, ACTIVE_r\}$$
$$P_{Starter} = \{tOut_r, wTick_r\}$$
$$T_{Starter} = \{$$
$$\quad (ACTIVE_r, true, \{\}, tOut_r, DISABLED_r),$$
$$\quad (DISABLED_r, true, \{\}, wTick_r, DISABLED_r)$$
$$\quad \}$$
$$L_{0_{Starter}} = \{ACTIVE_r\}$$

Figure 13 contains a graphical representation of the component. The $wTick$ is needed to ensure that the initial step is activated before any word is done.

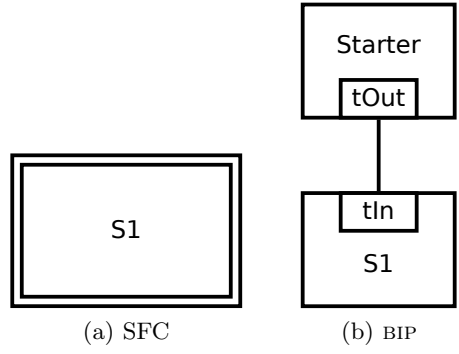

Figure 14: Initial Step

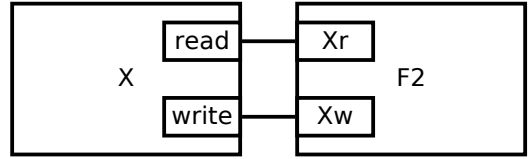

Figure 15: Use of Global Variable $X$ in F2

## 4.2 Transformation Steps

The following describes the steps necessary to convert an SFC program to a compound BIP component. All created instances of atomics are added to $A$.

For each global variable $x \in X$ we create an instance of the BIP Global Variable atomic component $\hat{x}$:

$$\hat{A}_X := \{\hat{x} := \text{CreateGV}(x) | x \in X\}$$

For each Action $a \in A$ create an instance $\hat{a}$ of it in BIP.

$$\hat{A}_A := \{\hat{a} := \text{CreateAction}(a) | a \in A\}$$

For each Action $a \in A$ create an instance of ACB component $\hat{b}_a$ in BIP and connect the *work* and *done* ports correspondingly (see Figure 16b).

$$\hat{A}_B := \left\{\hat{b}_a := \text{CreateACB}(a) | a \in A\right\}$$
$$\hat{C}_A := \left\{\left(\left(\hat{b}_a, work\right), \{(\hat{a}, work)\}\right) | \hat{a} \in \hat{A}_A\right\}$$
$$\cup \left\{\left(\left(\hat{b}_a, done\right), \{(\hat{a}, done)\}\right) | \hat{a} \in \hat{A}_A\right\}$$

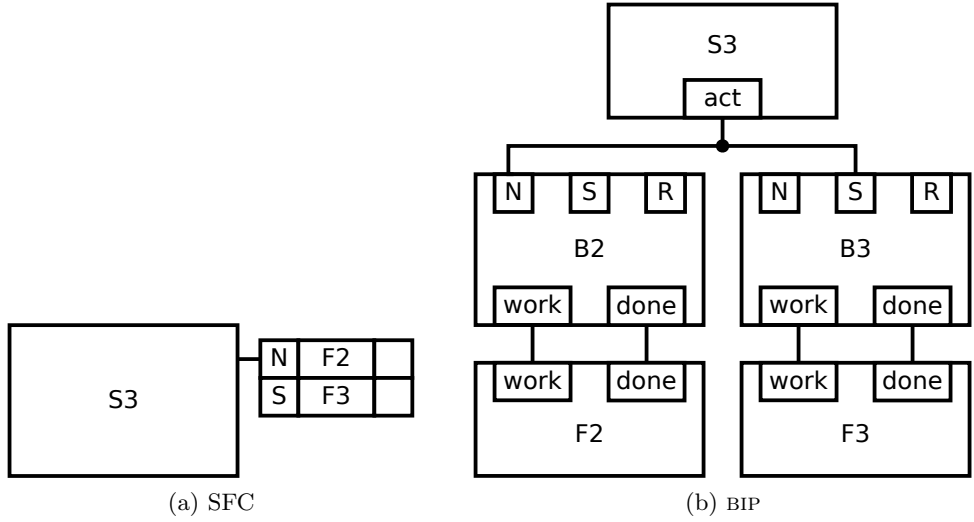

Figure 16: Action

For each SFC Step $s \in S$ instantiate the BIP Step component $\hat{s}$.

$$\hat{A}_S := \{\hat{s} := \text{CreateStep}(s) | s \in S\}$$

For each initial step $s \in S_0$ create a BIP Starter instance $\hat{r}_s$ and create a connector to $\hat{s}$ (Figure 14).

$$\hat{A}_{S_0} := \{\hat{r}_s := \text{CreateStarter}(s) | s \in S_0\}$$
$$\hat{C}_{S_0} := \{((\hat{t}_s, tOut), \{(\hat{s}, tIn)\}) | t_s \in A_{S_0}\}$$

For each SFC transition $(t_{src}, g, t_{dst}) = t \in T$ create a guard component $\hat{g}_t$ in BIP and connect it to $tOut$ of $t_{src}$ and $tIn$ of $t_{tgt}$. If the reading of variables is needed for evaluating the condition, necessary read ports, locations and transitions are created according to Figure 9.

$$\hat{A}_G := \{\hat{g}_t := \text{CreateStarter}(t) | t \in T\}$$
$$\hat{C}_T := \bigcup_{(t_{src}, t_g, t_{tgt}) = t \in T} \left\{ \left( (\hat{g}_t, guard), \bigcup_{s \in t_{src}} \{(\hat{s}, tOut)\} \cup \bigcup_{s \in t_{tgt}} \{(\hat{s}, tIn)\} \right) \right\}$$

Connect all actions and guards to global variables if necessary (see Figure 15).

$$\hat{C}_X := \bigcup_{(\hat{x}, \hat{a}) \in \hat{A}_X \times \hat{A}_A} \{((\hat{x}, read), \{(\hat{a}, \hat{x}_r)\}), ((\hat{a}, \hat{x}_w), \{(\hat{x}, write)\})\}$$
$$\cup \bigcup_{(\hat{x}, \hat{g}) \in \hat{A}_X \times \hat{A}_G} \{((\hat{x}, read), \{(\hat{a}, \hat{x}_r)\})\}$$

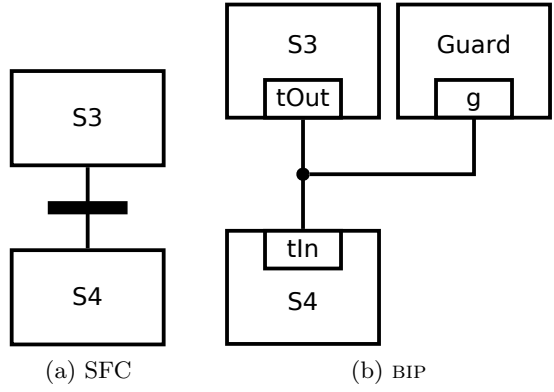

(a) SFC        (b) BIP

Figure 17: Transition

For each step $s \in S$ create a connector to the ACB $\hat{b}$ of all actions it activates. Use the port of the ACB according to the action qualifier. For the $P0/P1$ qualifier connect the $N$ port of the ACB to the guards $act$ port.

$$\hat{C}_B := \bigcup_{s \in S} \left\{ \left( (\hat{s}, act), \bigcup_{\substack{(a,q) \in s.\Omega \\ q \in \{N,S,R\}}} \left\{ \left( \hat{b}_a, q \right) \right\} \right) \right\}$$

$$\cup \bigcup_{(t_{src},t_g,t_{tgt})=t \in T} \left\{ \left( (\hat{g}_t, act), \bigcup_{\substack{(a,P0) \in s.\Omega \\ s \in t_{src}}} \left\{ \left( \hat{b}_a, N \right) \right\} \cup \bigcup_{\substack{(a,P1) \in s.\Omega \\ s \in t_{tgt}}} \left\{ \left( \hat{b}_a, N \right) \right\} \right) \right\}$$

Create an instance of the SFC manager component $\hat{m}$ and connect the $wTick$, $tTick$ and $fTick$ ports to all components that expect these signal.

$$\hat{C}_M := \{((\hat{m}, wTick), \{(\hat{a}, wTick) | \hat{a} \in \hat{A}_B \cup \hat{A}_X \cup \hat{A}_{S_0}\})\}$$
$$\cup \{((\hat{m}, tTick), \{(\hat{a}, tTick) | \hat{a} \in \hat{A}_B \cup \hat{A}_G \cup \hat{A}_X\})\}$$
$$\cup \{((\hat{m}, fTick), \{(\hat{a}, fTick) | \hat{a} \in \hat{A}_S \cup \hat{A}_G\})\}$$

Putting everything together

$$\hat{A} := \hat{A}_X \cup \hat{A}_A \cup \hat{A}_B \cup \hat{A}_S \cup \hat{A}_{S_0} \cup \hat{A}_G \cup \{\hat{m}\}$$
$$\hat{C} := \hat{C}_X \cup \hat{C}_A \cup \hat{C}_{S_0} \cup \hat{C}_T \cup \hat{C}_B \cup \hat{C}_M$$
$$\mathcal{B} := (\hat{A}, \hat{C})$$

In the following some samples for the transformation of transition are provided.

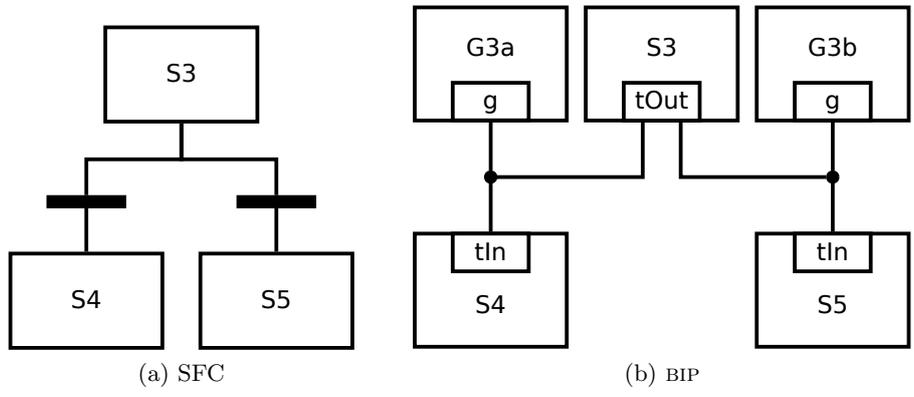

Figure 18: Divergence

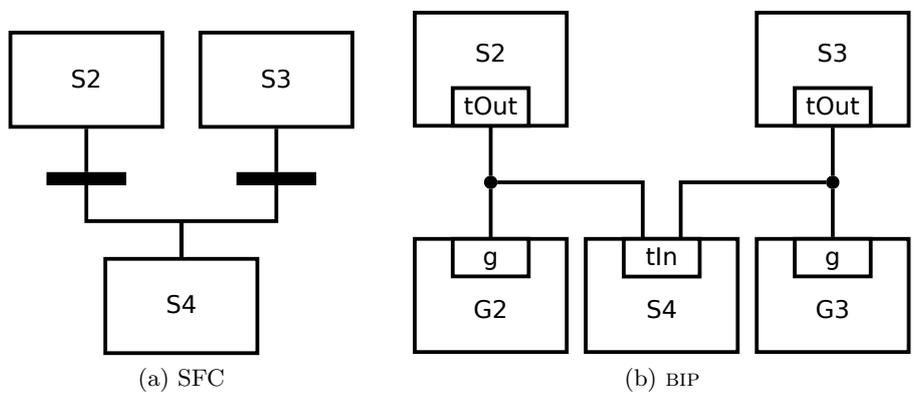

Figure 19: Convergence

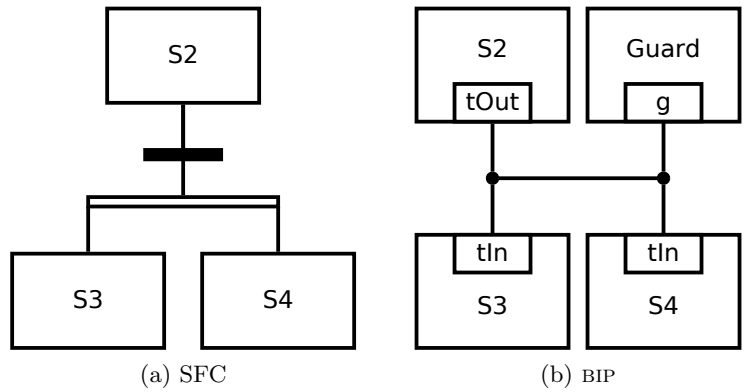

Figure 20: Parallel Divergence

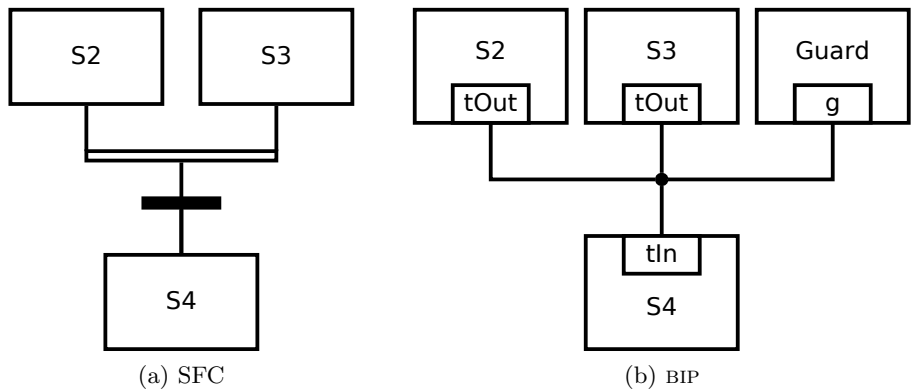

Figure 21: Parallel Convergence

Figure 17 shows a simple transition from one step to the following step. In BIP *tOut* of the source step and *tIn* of the destination step are connected together with a guard the checks the condition.

An alternative branch is shown in Figure 18. Here we have one connector and a single guard for each branch in BIP. Depending on which guard evaluates true first the associated following step is taken. Figure 19 shows the opposite. Two branches are united. In BIP each branch has one connection to *tIn* of the following step.

A parallel branch (Figure 20) looks similar to an normal transition, expect that it connects one source step with two following steps. It has also just one guard. The merge of parallel branches is shown in Figure 21. The connector can only switch, of all source steps are active and the guard evaluates to true.

# 5 Invariant Preservation of the SFC to BIP Transformation

In this section, we formally define and prove our invariant transformation function. We consider in this work only the non-extended SFCs. Given such an SFC $\mathcal{S} = (X, A, S, S_0, T, \sqsubset, \prec)$ and the transformed BIP model $\mathcal{B} = T(\mathcal{S})$, the invariant transformation function $T_I$ takes an invariant on a BIP model and returns an invariant on an SFC model. Our final goal is to prove that $\mathcal{B} \models \hat{I} \rightarrow \mathcal{S} \models T_I(\hat{I})$ holds. By definition, an invariant is a predicate on configurations that holds for all reachable configurations. The set of reachable configurations of SFC and BIP is defined in Sections 2.2 and 3.1, respectively.

The BIP model $\mathcal{B} = (\mathcal{A}, \mathcal{C})$ transformed from $\mathcal{S}$ contains a set of atomic components $\mathcal{A} = \hat{A}_X \cup \hat{A}_A \cup \hat{A}_B \cup \hat{A}_S \cup \hat{A}_{S_0} \cup \hat{A}_G \cup \{\hat{m}\}$. The system state (configuration) of any BIP model can be denoted as $\hat{c} = (atl, \sigma)$, where $atl : \mathcal{A} \rightarrow L$ is a function that returns the current location of each component and $\sigma$ is function that returns the current value of variables in each component. The notion $var(\hat{s})$ denotes the set of variables of component $\hat{s}$ and $\hat{s}.v$ denotes the particular variable $v$ of component $\hat{s}$.

**General Proof Sketch** We divide the proof of invariant preservation for model transformation into two steps. In the first step, we introduce a relation $R(c, \hat{c})$ between a reachable configuration $c$ in the original SFC model and a reachable configuration $\hat{c}$ in the transformed BIP model. Our first proof goal is to show that for each SFC transition $(c, c') \in [\![SM]\!]_{SFC}$, we can always find the corresponding BIP configuration pair $(\hat{c}, \hat{c}')$ such that $R(c, \hat{c}) \wedge R(c', \hat{c}') \wedge (\hat{c} \xrightarrow{\text{BM}}^{+}_{BIP} \hat{c}')$, where $\hat{c} \xrightarrow{\text{BM}}^{+}_{BIP} \hat{c}'$ is true if their exists a sequence of transitions contained in $[\![BM]\!]_{BIP}$ that transforms $\hat{c}$ to $\hat{c}'$. This property guarantees that the behavior in the SFC domain can always be traced in the BIP domain. Our proof goal in the second step is to show that the invariants are preserved for two configurations in the relation $R$.

Let's first assume that the two proof goals discussed above are true. Consider an arbitrary invariant $\hat{I}$ in BIP and the transformed predicate $T_I(\hat{I})$ in SFC domain, we want to show that $T_I(\hat{I})$ is indeed an invariant of the original SFC model. For that, we need to show that $T_I(\hat{I})$ holds on each reachable configuration of the SFC. According to the first proof goal, for an arbitrary reachable configuration $c$ in an SFC, we can find a configuration $\hat{c}$ in BIP such that $R(c, \hat{c})$ holds. Since an invariant must hold for all configurations by definition, $\hat{I}$ must also hold for $\hat{c}$. The second proof goal then states that the predicate $T_I(\hat{I})$ holds for $c$. Thus, we can conclude that for any invariants in BIP, the transformed predicate is an invariant of the corresponding SFC.

The relation $R$ can be formally defined as follows:

$$\forall c = (f, activeS, activeA), \forall \hat{c} = (atl, \sigma), \; R(c, \hat{c}) \; holds \; iff$$
$$rule_1(c, \hat{c}) \land rule_2(c, \hat{c}) \land rule_3(c, \hat{c}), where$$
$$rule_1(c, \hat{c}) = \forall \; s \in S, \hat{s} \in \hat{A}_S \; . \; s \in activeS \equiv \neg(atl(\hat{s}) = DISABLE)$$
$$rule_2(c, \hat{c}) = \forall \; a \in A, \hat{a} \in \hat{A}_A, \hat{b}_a \in \hat{A}_B \; . \; a \in activeA \equiv (\sigma(\hat{b}_a.e) = 1)$$
$$rule_3(c, \hat{c}) = \forall \; x \in X, \hat{x} \in \hat{A}_X \; . \; f(x) = \sigma(\hat{x}.v)$$

The first step of the proof is to shown that both $G1_a$ and $G1_b$ holds:

$$G1_a : R(c_0, \hat{c}_0)$$
$$G1_b : \forall c, c', \hat{c} \; . \; c \xrightarrow{SM}_{SFC} c' \land R(c, \hat{c})$$
$$\rightarrow \exists \hat{c}' \; . \; \hat{c} \xrightarrow{BM}^+_{BIP} \hat{c}' \land R(c', \hat{c}') \land BM = T(SM)$$

The second step of our proof is to show that for all configuration pairs in the relation, the invariant is preserved. Formally, the second proof goal $G2$ is:

$$G2 = \forall c, \hat{c} \; . \; R(c, \hat{c}) \; \rightarrow \; (\forall \hat{I} \; . \; \hat{c} \models \hat{I} \rightarrow c \models T_I(\hat{I}))$$

where $\hat{I}$ is a invariant in BIP model and $T_I$ is the invariant transformation function that maps $\hat{I}$ to an invariant in SFC domain. The notion $\hat{c} \models \hat{I}$ means $\hat{c}$ satisfies $\hat{I}$.

**Proof of $G1_a$** The proof that $G1_a = R(c_0, \hat{c}_0)$ holds can be done based on the model transformation rules:

- for each initially active step, a starter component is added in the BIP model, which triggers the step component to the $ACTIVE$ location ($rule_1$ fulfilled);

- no action should be active in $c_0$ and the initial value of variable $e$ in the ACB blocks are set to false ($rule_2$ fulfilled);

- the value of variable $v$ of GV components are initialized according to the SFC variables ($rule_3$ fulfilled).

Based on the above transformation rules, it can be trivially shown that the $R(c_0, \hat{c}_0)$ holds.

**Proof of $G1_b$** For the subgoal $G1_b$, we make a case distinction on three possible types of transitions of the SFC.

1. In an *executeAction* transition, some action $a$ is executed which updates the value of SFC variables. An action can be executed only if it is active, hence, $a \in activeA$. In the corresponding BIP model, the SFC manager enforces that all action execution is performed between the $fTick$ of the previous cycle and the $tTick$ of the current cycle. Let $\hat{a}$ be the action component and $\hat{b}_a$ be the ACB component created for $a$. Since the relation $R(c, \hat{c})$ holds, the $e$ variable in $\hat{b}_a$ is true according to $rule_2$. From the structure of ACB component (Figure 7), we can see that $\hat{b}_a$ can be on the $ENABLE$, $IDLE$ or $ACTIVE$ location. In all cases, the transition from location $ACTIVE$ to $WORK$ will be taken after next $wTick$ signal. This transition triggers the execution of action component $\hat{a}$, which performs the same set of operations as specified in $a$, because the model transformation just copies those operations. The transition also set the value of $\hat{b}_a.e$ to $false$. Execution sequence of the actions in the BIP model is guaranteed to be the same as in original SFC model, since transition from location $ACTIVE$ to $WORK$ in each ACB component is prioritized according to the order $\sqsubset$. Let $\hat{c}'$ be the BIP configuration in which the action $\hat{a}$ is just done ($\hat{b}_a$ in $WORKED$ location). It can be easily seen that $rule_2$ in the definition of relation $R$ holds for $(c', \hat{c}')$. On the other hand, since $tTick$ interaction has the lowest priority, no $tTick$ is issued during the transition from $\hat{c}$ to $\hat{c}'$. So, no step component goes to location $DISABLE$, which means $rule_1$ holds for $(c', \hat{c}')$. In addition, the relation $R(c, \hat{c})$ says that the current values of variable $v$ in GV components are the same as that of SFC variables $X$. Hence, the input of operations in $\hat{a}$ is the same as that of $a$. Since component $\hat{a}$ doesn't contain any internal variables, the output values will be the same, thus $rule_3$ holds for $R(c', \hat{c}')$. We can conclude that we reach an configuration $\hat{c}'$ such that $R(c', \hat{c}')$ holds.

2. In a $stepTransition$, the set of active steps is updated by taking some enabled transition $t$ and the rest of the configuration is not affected. According to the operational semantics of SFCs, step transitions are performed after execution of all active actions identified from the previous cycle. In the transformed BIP model, we achieve this point after the $tTick$ signal. At this moment, all ACB components are in $WAIT$ location and all guard components are ready to read values from $GV$ components. Let $\hat{t}_g$ be the guard component created for $t$. According to the relation $R(c, \hat{c})$, the variable values of $GV$ components are the same as SFC variables $X$. This guarantees that, if $t$ can be taken in the original SFC model, the condition $g$ in the $\hat{t}_g$ component must evaluate to $true$. During model transformation, the transitions from location $ACT$ to $DONE$ in all guard components are assigned to higher priority than $fTick$, hence, the interaction $((\hat{t}_g, guard), \{(\hat{s_1}, tOut), ..., (\hat{s_n}, tOut), (\hat{s_1}, tOut), ..., (\hat{s_m}, tOut)\})$ will be taken before next $fTick$ signal, where $\{\hat{s_1}, ...\hat{s_n}\}$ are BIP atomics created for steps $s \in t_{src}$ and $\{\hat{s_1}, ..., \hat{s_m}\}$ are the corresponding components for steps $s \in t_{tgt}$. This interaction triggers all source steps of transition $t$ to $DISABLE$ location and triggers all target steps of $t$ to $ACTIVE$ location. It can be easily shown that the $rule_1$ of relation $R$ is fulfilled after this interaction. On the other hand, in both SFC and BIP model, the set of active actions and the values of variables do not change. Thus we reach a configuration $\hat{c}'$ such that $R(c', \hat{c}')$ holds.

3. The third type of SFC configuration transition is the computation of active actions, which is the last phase in an SFC execution cycle. In the transformed BIP model, we reach this point after an $fTick$ interaction. The relation $R(c, \hat{c})$ says that for each active step in the SFC, the corresponding step component in BIP must not in $DISABLE$ location. Hence, before the next $wTick$, all active steps components will activate the associated actions (Figure 8), since $wTick$ interaction has the lowest priority. According to the model transformation rules, for each action contained in the set $\Omega$, a connector is constructed that connects the activation port of the step component with the ACB component. Thus, the set of active actions will be the same in both SFC and BIP models, given the set of active steps are the same. This guaranteed that $rule_2$ holds. Since the rest part of the configuration doesn't change in both models, we reach a configuration $\hat{c}'$ such that $R(c', \hat{c}')$ holds.

As shown above, for all types of configuration transitions in SFC, the proof goal $G1_b$ holds. Note that we made the implicit assumption in our proof that we either use the same datatypes for both SFC and BIP model or no overflows and underflows of value ranges occur.

Before we present a proof of $G2$, a formalism for describing the invariants in SFCs and BIP models is needed, which is presented in the following paragraphs.

**Invariants on SFC** The syntax of an SFC invariant can be expressed as follows:

$$
\begin{aligned}
I &::= i \wedge i \\
i &::= i' \\
i' &::= i' \vee i' \\
i' &::= i'' \\
i'' &::= \neg i'' \\
i'' &::= C \mid AS \mid AA \\
C &::= cond(X) \\
AS &::= s \in activeS \\
AA &::= a \in activeA
\end{aligned}
$$

Where $cond(X)$ is a predicate on the set of SFC variables $X$. For an SFC, we can statically analyze the model and find the activation relation between steps and actions, which can help us to express the invariants in SFC models. Let $S_N(a) = \{s \in S \mid a \in s.\Omega\}$ denote the set of steps associated with an action $a$. A structural invariant that holds for all non-extended SFCs is:

$$\left(\left(\bigvee_{s \in S_N(a)} s \in activeS\right) \wedge a \in activeA\right) \vee \left(\left(\bigwedge_{s \in S_N(a)} \neg(s \in activeS)\right) \wedge \neg(a \in activeA)\right)$$

This invariant says if at least one step that can invoke the action $a$ is in active state, the action $a$ is also in active state, otherwise $a$ is not active.

**Invariants on BIP Models** An invariant in a BIP model can be expressed using the following syntax:

$$\hat{I} ::= i$$
$$i ::= i \wedge i$$
$$i ::= i'$$
$$i' ::= i' \vee i'$$
$$i' ::= i''$$
$$i'' ::= \neg i''$$
$$i'' ::= C \mid L \mid C \wedge L$$
$$C ::= cond(var(\hat{s}))$$
$$L ::= atl(\hat{s}) = l$$

where $atl(\hat{s}) = l$ is true if the component $\hat{s}$ is at location $l$, and $cond(var(\hat{s}))$ is a predicate on the variables of $\hat{s}$. The structure of this invariant language reflects the invariant generated by D-Finder [5]. Some examples are given in Section 3.2.

**Invariant Transformation** From the previous definition, we see that any BIP invariant can be written as logical expressions of two basic sets of elements: predicates on variables of individual component and predicates on locations of individual components. We introduce an invariant transformation function $T_I$, which takes an BIP invariant $\hat{I}$ and returns the corresponding invariant in the SFC domain. It is defined inductively:

$$T_I(\hat{I}_1 \wedge \hat{I}_2) = T_I(\hat{I}_1) \wedge T_I(\hat{I}_2)$$
$$T_I(\hat{I}_1 \vee \hat{I}_2) = T_I(\hat{I}_1) \vee T_I(\hat{I}_2)$$
$$...$$

As it can be seen, the function $T_I$ keeps the structure of BIP invariants and maps each elementary BIP predicate to a corresponding predicate in the SFC domain. Given a BIP model $\mathcal{B} = T(\mathcal{S})$ transformed from an SFC model $\mathcal{S}$, the mapping of elementary predicates on $\mathcal{B}$ to predicates on $\mathcal{S}$ is described using following rules. The notion $\hat{s}.L$ denotes the set of all locations of BIP atomic $\hat{s}$.

1. For a predicate on variables $p = cond(var(\hat{s}))$ we distinguish the following cases:
    - if $\hat{s}$ is a GV component created for SFC variable $x$, $\hat{s}$ has only two variables $v$ and $t$ by definition. Since the relationship between $v$ and $t$ is a simple assignment, the condition can be written in the form $cond_t(t) \wedge cond_v(v)$. Then, the corresponding predicate in SFC domain is $T_I(p) = cond_v(x)$, i.e. we ignore the condition on $t$ and apply the condition on $v$ to SFC variable $x$;

- if $\hat{s}$ is a ACB component, it contains a boolean variable $e$. The predicate $p$ is a boolean expression on these variables. Let $a$ be the corresponding SFC action that $\hat{s}$ is created for, the transformation of $p$ is done by keeping the structure of expression and doing the replacement: $T_I(e) = a \in activeA$;

- if $\hat{s}$ is an other component, $T_I(p) = true$. Actually, this rule is not used, since only GV and ACB component contain variables according to the model transformation rules.

2. For predicates on locations we distinguish the following cases:
    - if $\hat{s}$ is a step component, $T_I(atl(\hat{s}) = DISABLE) = \neg(s \in activeS)$ and $T_I(atl(\hat{s}) = l) = s \in activeS, \forall l \in \hat{s}.L \backslash \{DISABLE\}$, where $s$ is the step in SFC that $\hat{s}$ is transformed from;
    - if $\hat{s}$ is a ACB component created for SFC action $a$, $T_I(atl(\hat{s}) = ENABLE) = \bigvee_{s \in S_N(a)} s \in activeS$ and $T_I(atl(\hat{s}) = l) = false, \forall l \in \hat{s}.L \backslash \{ENABLE\}$;
    - if $\hat{s}$ is other component, $T_I(p) = false$.

In the invariant transformation procedure, some unused predicates on locations are eliminated by setting them to false in the transformed SFC invariant. This is safe because the BIP component invariants have a disjunction form as mentioned before.

**Example.** For an ACB component $\hat{s}$ (Figure 7), an obvious invariant can be found as: $\hat{I} = \neg(atl(\hat{s}) = ENABLE) \vee (atl(\hat{s}) = ENABLE \wedge e)$. By applying the transformation rules, we can obtain $I = (\neg \bigvee_{s \in S_N(a)} s \in activeS) \vee ((\bigvee_{s \in S_N(a)} s \in activeS) \wedge a \in activeA)$,
which is an invariant contained in the example SFC structural invariant we found before.

Another set of invariants that can be identified in BIP models are interaction invariants. Let's consider a BIP model transformed from a simple SFC shown in Figure 17. The transformed BIP model consists of (besides other necessary components) two step atomics and one guard atomic. A connector $\{(S3, tOut), (S4, tIn), (Guard, g)\}$ is built by the transformation. The interaction described by the above connector is enabled if $atl(S3) = ACTIVE \wedge atl(S4) = DISABLE \wedge atl(Guard) = ACT \wedge Guard.g$ hold. An obvious interaction invariant that can be observed is that one of $S3$ and $S4$ must be in $DISABLE$ location, i.e. $atl(S3) = DISABLE \vee atl(S4) = DISABLE$ must hold. By applying the transformation rules, the predicate in SFC domain is $\neg S3 \in activeS \vee \neg S4 \in activeS$, which obviously holds by the semantics of SFC.

**Proof of** $G2$  The proof of second goal comes strait forwardly from the definitions of $R$ and $T_I$. Consider a configuration $c$ of an SFC model and a configuration $\hat{c}$ of the transformed BIP model, we now show that if $R(c, \hat{c})$ holds, then for an arbitrary invariant $\hat{I} \models \hat{c}$ in the BIP domain, the transformed predicate $I = T_I(\hat{I})$ holds for $c$ in the SFC domain. Since $I$ and $\hat{I}$ have the same structure, it is sufficient to show that all the elementary predicates in $\hat{I}$ evaluates to the same value as the transformed predicates in $I$. We make a case distinction on the two types of predicates. The first type are predicates on variables of some component $\hat{s}$. We can see from the model transformation rules only the $GV$ and $ACB$ component contains internal variables. The $rule_2$ and $rule_3$ in the definition of $R$ guarantees that the predicates before and after transformation $T_I$ evaluate to the same result. The second type are predicates on locations. As stated in the definition of $T_I$, only the predicates on step components and ACB components are kept. For the case of step components, the $rule_1$ guarantees the correctness. For the case of ACB components, the correctness comes from the structure of the ACB: when we move to the $ENABLE$ location, at least one step component has interacted with the ACB via the activation port, which means at least one step component must be active in the first place.

# 6 Transformation of SFC Safety Requirements into BIP Invariants

The BIP framework provides a set of invariant-based formal verification tools. An important usage scenario is using these existing tools to verify safety properties on the BIP level. Invariants discovered on BIP to show the absence of deadlocks fall into this class of properties. However, some safety requirements may already be specified on the original SFC models. For that, besides transforming the SFC model into BIP, we also need to translate these requirements into BIP domain.

A safety requirement specifies typically a set of non-safe configurations in the SFC $\mathcal{S}$ that should be never be reached. Here we provide a solution to ensure the unreachability of unsafe configurations. We define a translation function. It has the form $T_R : \mathcal{C} \to \mathcal{P}$ and takes an SFC configuration $c \in \mathcal{C}$ and returns a predicate $p \in \mathcal{P}$ on the transformed BIP model $\mathcal{B}$. Since the BIP model specifies more behavior, a translated predicate are not necessarily a single BIP configuration but can be a set of configurations.

This property translation function $T_R$ can be defined as follows:

$$T_R((activeS, activeA, f)) =$$
$$\forall s \in activeS \ . \ \neg(atl(\hat{s}) = DISABLE)$$
$$\land \forall a \in activeA \ . \ \hat{b}_a.e = 1$$
$$\land \forall x \in X \ . \ \sigma(\hat{x}.v) = f(x)$$

After translating the safety requirements, one has to ensure that it does indeed hold in the BIP domain. Verifying that such a safety requirement is fulfilled is equivalent to showing $\mathcal{B} \models \neg p$ where $p$ is a predicate specifying the unsafe configuration. This means $\neg p$ is an invariant of the system.

The transformation function stated above is safe because the returned invariants in the BIP domain are at most as strong as the corresponding invariants in the SFC domain, as discussed in Section 5.

## 7 Concluding Remarks

In this report we presented the formalization of the transformation from the SFC language of the IEC 61131–3 standard into BIP. Furthermore, we have presented transformations for invariants over IEC 61131–3 to invariants over BIP and vice versa. These transformations are accompanied by a correctness proof which is based on the semantics of the involved language.

The drawbacks of this work comprises the fact that our proof only deals with abstract transformation rules. We did not achieve a verified implementation yet. We are currently implementing the transformations in Java using the Eclipse Modeling Framework [2]. Regarding correctness guarantees, we are working on a certificate generation and checking infrastructure similar to those described in [6, 7, 8].

---

[2]*see* http://www.eclipse.org/modeling/emf/